\documentclass{article}
\usepackage{amsmath}
\usepackage{harvard}

\newtheorem{theorem}{Theorem}
\newtheorem{acknowledgement}[theorem]{Acknowledgement}

\input{tcilatex}
\begin{document}

\title{Quantum problem solving as simultaneous computation}
\author{Giuseppe Castagnoli \\
Pieve Ligure (Genova), giuseppe.castagnoli@gmail.com}
\maketitle

\begin{abstract}
I provide an alternative way of seeing quantum computation. First, I
describe an idealized classical problem solving machine that, thanks to a
many body interaction, reversibly and nondeterministically produces the
solution of the problem under the simultaneous influence of all the problem
constraints. This requires a perfectly accurate, rigid, and reversible
relation between the coordinates of the machine parts -- the machine can be
considered the many body generalization of another perfect machine, the
bounching ball model of reversible computation. The mathematical description
of the machine, as it is, is applicable to quantum problem solving, an
extension of the quantum algorithms that comprises the physical
representation of the problem-solution interdependence. The perfect relation
between the coordinates of the machine parts is transferred to the
populations of the reduced density operators of the parts of the computer
register. The solution of the problem is reversibly and nondeterministically
produced under the simultaneous influence of the state before measurement
and the quantum principle. At the light of the present notion of \textit{%
simultaneous computation}, the quantum speed up turns out to be
"precognition" of the solution, namely the reduction of the initial
ignorance of the solution due to backdating, to before running the
algorithm, a time-symmetric part of the state vector reduction on the
solution; as such, it is bounded by state vector reduction through an
entropic inequality.

\end{abstract}

\section{Introduction}

This work is about the notion of \textit{simultaneous computation}, a
fundamental computation mechanism implicit in quantum computation. It can
also be seen as an application of the notion of Gestalt to computation. The
following is a summary by section.

Section 2. I review the definition of Gestalt as truly simultaneous
dependence between all the quantitative variables describing a physical
situation.

Section 3. A perfect simultaneous dependence between all computational
variables enables a simultaneous form of computation in the idealized
classical framework. The problem addressed is solving a simultaneous system
of Boolean equations. The Boolean variables are mapped by real non-negative
variables (coordinates of the machine parts) submitted to idealized physical
constraints (perfectly accurate, rigid and reversible) representing the
simultaneous system of Boolean equations. The solution of the problem is
reversibly and nondeterministically produced under the simultaneous
influence of all equations -- in fact through an idealized many body
interaction. This machine can be seen as the many body generalization of
another perfect machine, the bounching ball model of reversible computation.

Section 4. Simultaneous computation turns out to be a representation of
quantum problem solving, an extension of the quantum algorithms that
comprises the physical representation of the problem-solution
interdependence. The simultaneous dependence between the coordinates of the
machine parts is transferred to the populations of the reduced density
operators of the parts of the computer register. Simultaneous dependence
becomes quantum correlation. Infinite precision is replaced by quantization
(Finkelstein, 2007). The solution of the problem is reversibly and
nondeterministically produced under the simultaneous influence of the state
before measurement and the quantum principle.

Section 5. The quantum speed up turns out to be "precognition of the
solution" -- the reduction of the initial ignorance of the solution due to
backdating, to before running the algorithm, a time-symmetric part of the
state vector reduction on the solution; as such, it is bounded by state
vector reduction through an entropic inequality.

Section 6. The notion of simultaneous computation is positioned within the
development of quantum computation.

\section{The notion of Gestalt}

In the definition of the Vienna circle, Gestalt is an organized coherent
whole whose parts are determined by laws intrinsic to the whole rather than
being independently juxtaposed or associated. Moreover, if the whole can be
described by quantitative variables, Gestalt is defined as truly
simultaneous dependence between all variables (e. g., Mulligan and Smith,
1988).

It can be argued that "truly simultaneous dependence" is only between the
variables related by a fundamental physical law. An example in classical
physics is Newton's law in the idealized case of a point mass: it
establishes a truly simultaneous dependence between force, mass, and
acceleration since the change of any one variable is correlated with an
identical change of the product, or ratio, of the other two. In view of what
will follow, I should remark that simultaneous dependence between variables
\ is \textit{mutual} , like correlation, and can be \textit{non-functional},
in the sense that the perturbation of any one variable does not univocally
"propagate" to the others. This is the case of Newton's law that,
correspondingly, can host in principle a nondeterministic many body
interaction.

Identified with the notion of physical law, Gestalt becomes a revisitation
of the Platonic notion of Form, or Idea (the Greek word Eidos translates
into Form, Idea, or Vision): "Ideas are objective perfections that exist in
themselves and for themselves, at the same time they are the cause of
natural phenomena, they keep phenomena bound together and constitute their
unity.

The concept of objectively perfect simultaneous dependence can be clarified
by resorting to the concept of mechanism. A fundamental law should be seen
as a perfect mechanism (perfectly accurate, rigid, and reversible) whose
degrees of freedom are the continuous variables related by the law -- it is
not the case that this mechanism gets deformed because of flexibility or
jams because of friction or irregularities. As we will see, an
underconstrained perfect mechanism, characterized by a non-functional
relation between its degrees of freedom, can host a nondeterministic many
body interaction.

\section{Simultaneous computation in an idealized classical framework}

By using the definition of Gestalt as absolutely simultaneous dependence
between all the quantitative variables describing a physical situation, one
can see in the first place that there is no Gestalt in classical
computation. Let us consider the idealized bouncing ball model of reversible
computation (Fredkin and Toffoli, 1982). The variables at stake are the
positions and momenta of each and every ball. Outside collisions, there is
no simultaneous dependence between the variables of different balls, which
are independent of each other. In the instant of (idealized) collision,
there is simultaneous dependence between the variables of the colliding
balls, but this is limited to ball pairs (there can be several collisions at
the same time, but involving independent ball pairs, with no simultaneous
dependence between the variables of different pairs). The simultaneous
collision between more than two balls is avoided since it would introduce
the many body problem.

This lack of Gestalt raises the question whether there is a form of
computation endowed with it. By assuming a perfect simultaneous dependence
between all computational variables, one can devise an idealized classical
machine that -- thanks to a many body interaction -- nondeterministically
produces the solution of a system of Boolean equations under the
simultaneous influence of all equations (the same simultaneous dependence
will represent quantum correlation).

We can start with the simplest "problem" represented by a single
unconstrained Boolean variable $x$ -- we will also use the auxiliary
variable $y=\overline{x}$. The problem solutions are of course $x=0$, $y=1$
and $x=1$,$~y=0$. Let $X,Y,Q$ be real non-negative variables. The Boolean
problem can be transformed into the problem of finding the solutions, for$%
~Q>0$, of the simultaneous equations

$\bigskip $%
\begin{equation}
\frac{X}{Q}+\frac{Y}{Q}=1,  \label{linear}
\end{equation}

\begin{equation}
\left( \frac{X}{Q}\right) ^{2}+~\left( \frac{Y}{Q}\right) ^{2}=1.
\label{nonlinear}
\end{equation}%
$Q=0$ implies $X=Y=0$. With $Q>0$, $\frac{X}{Q}\equiv x$ and$~\frac{Y}{Q}%
\equiv y$: one can see that $X=0,$ $Y>0~$corresponds to the Boolean values $%
x=0,~y=1~$and $X>0,$ $Y=0~$to $x=1,\ y=0$.

In this real variable representation of the Boolean problem, the solutions
can be computed by a many body interaction, as follows. Equation (\ref%
{linear}) can be represented by an idealized hydraulic circuit where $Q$ is
the coordinate of a piston feeding in parallel (through an incompressible
fluid) two pistons of even section and mass, and coordinates respectively $X$
and $Y$. Equation (\ref{nonlinear}) is represented by a differential
mechanism with non-linear (parabolic) cams applying to pistons $X$, $Y$, and 
$Q~$(I use the same symbol to denote the piston and its coordinate). The
initial configuration of the machine is $X=Y=Q=0$; it can be argued that any
movement of piston $Q$ from $Q=0~$to$~Q>0$ instantly produces a solution in
a nondeterministic way. This motion could be obtained by applying a force to
piston $Q$, then there would be no reason that either $X$ or $Y$ (in a
mutually exclusive way) move with $Q$, as either movement offers zero static
resistance to the force (there is only the inertia of the pistons). This
reversible, nondeterministic many body interaction should be postulated in
the present idealized classical framework, in the quantum framework it
becomes a representation of measurement. I should note that simultaneous
dependence is nonlinear in the nondeterministic transition from $Q=0~$to$%
~Q>0 $, and linear in the deterministic movement of the pistons in the
interval $Q>0$ -- in fact, with $Q>0~$and, say, $X=0$, the two equations (%
\ref{linear}) and (\ref{nonlinear}) make a redundant linear system in the
positive interval.

Unlike deterministic reversible processes, the present process is not
invertible -- in general one cannot go back and forth along the same
process. For example, we can think of connecting the input piston to an
ideal spring charged when $Q=0$. On the one side, there would be
oscillations without dissipation. On the other, at each oscillation, the
movement of the input piston from $Q=0~$to$~Q>0$ would randomly drag either $%
X~$or$~Y~$in a mutually exclusive way. \ 

This idealized computation mechanism can solve any system of Boolean
equations, namely of $N~$NAND equations $x_{i,3}=\func{NAND}%
(x_{i,1},x_{i,2}) $, with$~i=1,~...,~N$ and $x_{i,j}=x_{h,k}$ for some
assignements of $i,~j,~h,~k$. The hydraulic circuit becomes the series of an
input branch/piston $Q$ and $N$ quadruples of parallel branches/pistons $%
X_{i,j},~~j=1,~...,~4$. The four branches/pistons of each quadruple are
labeled by the Boolean values that satisfy the corresponding NAND equation.
For example, branches/pistons $X_{i,1};~X_{i,2};~X_{i,3};\ X_{i,4~}$ are
labeled by, respectively, $%
x_{i,1}=0,~x_{i,2}=0,~x_{i,3}=1;~x_{i,1}=0,~x_{i,2}=1,~x_{i,3}=1;~x_{i,1}=1,~x_{i,2}=0,~x_{i,3}=1;~x_{i,1}=1,~x_{i,2}=1,~x_{i,3}=0 
$. "Fluxes" $X_{i,j}$ in the branches of the same quadruple are made to be
mutually exclusive with one another by nonlinear transmissions between the
corresponding pistons and the total flux across branches labeled by the same
value of the same Boolean variable is made to be conserved across different
quadruples by linear transmissions between the corresponding pistons. By
applying a force to the input piston $Q$, the machine's motion from $Q=0~$to$%
~Q>0$ instantly produces a solution under the simultaneous influence of all
the problem constraints (in each quadruple, there is only one branch with
flux $>0$, the series of all these branches is labeled by a Boolean
assignment that solves the system). By using the partial OR (POR) gate
instead of the NAND gate, quadruples can be replaced by triples.

This idealized machine has the only purpose of introducing the idea of
simultaneous computation, namely of a computation mechanism that, thanks to
a perfect simultaneous dependence between many continuous computational
variables, nondeterministically produces the solution of a problem under the
simultaneous influence of all the problem constraints.

\section{Quantum computation as simultaneous computation}

To see that quantum computation is simultaneous computation, we should
replace the configuration space of the idealized classical machine by the
phase space of the quantum machine. Simultaneous dependence between the
coordinates of the machine parts becomes simultaneous dependence between the
populations of the reduced density operators of the parts of the computer
register. Let us consider for example the simplest problem of section 3, of
finding the solutions of a single unconstrained Boolean variable $x$. The
motion of the idealized machine from $Q=0$ to $Q>0$ is analogous to
measuring two qubits in the entangled state $\left\vert 0\right\rangle
_{X}\left\vert 1\right\rangle _{Y}+$ $\left\vert 1\right\rangle
_{X}\left\vert 0\right\rangle _{Y}$. Let us represent the populations of the
reduced density operator of the first (second) qubit by the variables $%
x_{11},x_{22}$ ($y_{11},~y_{22}$). The state before measurement corresponds
to the assignement $x_{11}=~x_{22}=y_{11}=~y_{22}=\frac{1}{2}$, the state
after measurement to $x_{11}=1,$ $x_{22}=0,$ $y_{11}=0,$ $y_{22}=1~$or, in a
mutually exclusive way, $x_{11}=0,~x_{22}=1,~y_{11}=1,~y_{22}=0$. The
correspondence between the coordinates of the idealized classical machine
and the populations of the quantum machine is:

\begin{equation}
\frac{X}{Q}=x_{11}=1-x_{22},~\frac{Y}{Q}=y_{11}=1-y_{22}.
\label{correspondence}
\end{equation}

\bigskip The transition imposed by the quantum principle is isomorphic with
the transition from $Q=0~$to$\ Q>0$\ of section 3 and can be represented in
exactly the same way by adding, to equations (\ref{correspondence}),
equations (\ref{linear}) and (\ref{nonlinear}), repeated here for
convenience:

\begin{equation}
\frac{X}{Q}+\frac{Y}{Q}=1,  \label{poplinear}
\end{equation}
\begin{equation}
\left( \frac{X}{Q}\right) ^{2}+\left( \frac{Y}{Q}\right) ^{2}=1,
\label{popnonlinear}
\end{equation}%
We can see that the infinite precision required by the classical machine is
replaced by quantization (Finkelstein, 2007).

More in general, the nondeterministic production of the solution under the
simultaneous influence of all the problem constraints of the idealized
classical model becomes the nondeterministic production of the solution
under the joint influence of the state before measurement and the condition
established by the quantum principle, that the state before measurement is
projected on the eigenspace of one eigenvalue of the measured observable,
with probability the square of the modulus of the corresponding amplitude.
State vector reduction performs the computation (symbol handling) that
transforms the symbolic description of the state before measurement into the
symbolic description of the state after measurement. The computation is
performed by simultaneously solving a system of Boolean equations -- not by
the causal propagation of an input into an output (Castagnoli, 1999).

I should note that simultaneous dependence encompasses the whole sequence
preparation/unitary evolution/quantum measurement. In fact there is
simultaneous dependence between any two populations at any two times $%
t_{1},~t_{2}$\ along the sequence, as follows. As before, populations should
be seen as variables that switch from an assignement to another in
correspondence of state vector reduction. We can take as independent
variables the amplitudes $\left\{ \alpha _{i}\right\} ,~i=1,2,...,N,$ of the
basis vectors in the preparation of the quantum system. During unitary
evolution, the amplitudes of the basis vectors evolve into linear
combinations $f_{i,t}\left( \left\{ \alpha _{i}\right\} \right) $ of the
independent variables. Setting the independent variables to the values they
have in the preparation yields the usual quantum evolution (the forward
evolution). Quantum measurement changes this evolution into the backward
evolution, which undergoes the same unitary transformation but starts with a
different preparation and ends with the state after measurement. There is
simultaneous dependence between any two populations at any two times $%
t_{1},~t_{2}$ since the change of one population from the forward to the
backward value (in correspondence with the transition from $Q=0~$to $Q>0$)
is correlated to the change of the other. Simultaneous dependence is like
between the polarizations of two photons in a polarization entangled state;
the determination of one polarization determines the other, also at a
previous or subsequent time along the sequence preparation/unitary
evolution/measurement, and vice-versa.

\section{An explanation of the speed up}

Simultaneous computation offers a way of explaining the speed up of the
quantum algorithms. In the years 1997-2000 I published, with others, a few
papers that ascribe the quantum speed up to the non causal
joint-determination of the measurement outcome by the state before
measurement and the quantum principle (Castagnoli 1997, 1999, Castagnoli et
al., 2000, Castagnoli and Finkelstein, 2000). We showed that the
joint-determination associated with entanglement and disentanglement, the
former due to quantum parallel computation, the latter to quantum
measurement, was responsible for all the speed-ups discovered until then.
Soon afterwards (Raussendorf and Briegel, 2000) there was the first paper on
cluster computing, where the use of entanglement and disentanglement by
quantum measurement becomes explicit. The notion propounded in the
following, that the speed up depends on backdated state vector reduction and
is therefore entropically bounded by state vector reduction, is new,
although already implicit in (Castagnoli and Finkelstein, 2000).

We have seen that, in the quantum context, simultaneous computation becomes
the nondeterministic production of the solution under the joint influence of
the quantum principle and the state before measurement. This is evident in
the algorithms of Simon (1994) and Shor (1994). Here "function evaluation"
produces an entangled state of the form $\sum_{x=1}^{N}\left\vert
x\right\rangle _{X}\left\vert f\left( x\right) \right\rangle _{F}$, with $%
f\left( x\right) $ a periodic function of $x$. The final measurement is
equivalent to measuring the value of the function immediately after function
evaluation, for the retroactivity of state vector reduction in a reversible
evolution. Joint-determination extradynamically filters, out of an
exponential number of arguments, all the arguments corresponding to a common
value of the function. By applying the quantum Fourier transform to the
superposition of such arguments, one extracts the period of the function.

On the contrary, joint-determination is completely hidden in Deutsch's
(1985) and Grover's (1996) algorithms, which yield their speed ups through
unitary evolutions -- I am presently considering Cleve's et al. (1997)
revisitation of Deutsch's algorithm and Grover's algorithm for a database
size that provides no probability of error. Apparently, there is no
entanglement between computer registers and no nondeterministic production
of the solution. However, this can be ascribed to the fact that these
algorithms physically represent only the procedure that leads to the
solution, whereas the interdependence between the problem and the solution
is disregarded.

We are dealing with quantum games. One player chooses at random one of the
four functions in Deutsch's problem, or a data base location in Grover's
problem, the other player must find out the choice of the first player (a
character thereof in Deutsch's problem), but the physical representation
does not comprise the random generation of the move of the first player.\
Simultaneous computation (problem-solution interdependence) does not appear
since the problem is not represented physically.

Let us focus on Grover's algorithm and let the size of the database be $N$.
In the conventional algorithm, the quantum database is represented by a
quantum computer that, given an input $x$, computes $\delta \left(
k,x\right) $, where $\delta $ is the Kronecker function and $k$ is the
database location randomly chosen by the first player. For each input $x$
provided by the second player, the computation of $\delta \left( k,x\right) $
tells whether it is the database location chosen by the first player. The
second player prepares the input register $X$ in an even superposition of
all the possible values of $x$. To find out the choice of the first player,
the algorithm has to compute $\delta \left( k,x\right) $ the order of $\sqrt{%
N}$ times, instead of $N$ like in the classical case.

To physically represent the problem, it suffices to represent the random
generation of $k$ on the part of the first player -- then the computation of 
$\delta \left( k,x\right) ~$copes for problem-solution interdependence. To
this end, we add an ancillary register $K~$prepared in a superposition of
all the possible values of $k$. The extended algorithm repeatedly computes $%
\delta \left( k,x\right) $ as before but now for a superposition of all the
possible combinations of values of $k$ and $x$. This entangles each possible
value of $k$ with the corresponding solution (the same value of $k$) found
by the second player at the end of the algorithm. For example, with database
size $N=4$, the state before measurement is:

\begin{equation}
\frac{1}{2\sqrt{2}}\left( \left\vert 00\right\rangle _{K}\left\vert
00\right\rangle _{X}+\left\vert 01\right\rangle _{K}\left\vert
01\right\rangle _{X}+\left\vert 10\right\rangle _{K}\left\vert
10\right\rangle _{X}+\left\vert 11\right\rangle _{K}\left\vert
11\right\rangle _{X}\right) \left( \left\vert 0\right\rangle _{F}-\left\vert
1\right\rangle _{F}\right)  \label{gstate}
\end{equation}

Measuring the content of registers $K~$and $X$ determines the moves of both
players -- also representing the random choice of the value of $k~$on the
part of the first player. The content of register $K$ is known to the first
player, that of register $X$ to the second player. The state vector
reduction induced by measuring the content of register $K$ can be backdated
to before running the algorithm. This leaves the initial preparation of
register $X$ -- a superposition of all the possible values of $x$ --
unaltered (because of the unitary transformations in between) and brings
that of register $K$ to a sharp value, thus representing exactly the
original Grover's algorithm. In view of what will follow, I should note that
the state of the computer register also represent the state of knowledge of
the value of $k$ on the part of the second player. The initial state of the
register -- where $K$ is in an even superposition of all the possible values
of $k$ -- represents ignorance of the value of $k$. The fully entangled
state (\ref{gstate}) at the end of the algorithm, as well as the outcome of
final measurement, represents knowledge of it.

Thus, by completing the physical representation of Grover's algorithm, one
finds again a succession of entanglement and disentanglement, and
simultaneous computation through joint-determination of the measurement
outcome by the state before measurement and the quantum principle.
Joint-determination can be seen as mutual determination between the contents
of the two entangled registers $K$ and $X$, which justifies the square root
speed up with respect to a classical database search, where the content of
the former register determines that of the latter and not vice-versa. By
ascribing the speed up to mutual determination between register contents,
one finds that it is bounded by state vector reduction through an entropic
inequality, as follows.

Mutual determination is symmetrical, it can be represented by saying that
the contents of the two registers are determined by the measurement of the
first (second) bit of register $K~$and the second (first) bit of register $X$%
. Thus Grover's algorithm is equivalent to the following game. We should
think to arrange the $N$ database locations in a matrix of $\sqrt{N}$\
columns and $\sqrt{N}$ rows -- with $N=4$ the row can be identified by the
first bit of either register, the column by the second bit. At the end of
Grover's algorithm, the first player determines, say, the row by measuring
the first bit of register $K$ in state (\ref{gstate}). This is equivalent to
determining the row before running the algorithm, for what said before. The
second player determines (and knows) the column by measuring the second bit
of register $X$. The related state vector reduction can be backdated to
before running the algorithm, namely to the initial preparation of the two
registers $K$ and $X$, each in an even superposition of all the possible
values of, respectively, $k$ and $x$. This leaves the initial preparation of
register $X$ unaltered (because of the unitary transformations in between)
and reduces that of register $K$ to the superposition of all the values of $%
k $ ending by that bit (determining the column before running the
algorithm). In this picture, Grover's algorithm searches just the row
randomly chosen by the first player, which justifies the $O\left( \sqrt{N}%
\right) $ computations of $\delta \left( k,x\right) $, i. e. the square root
speed up (of course the picture should be symmetrized for the exchange of
columns and rows).

The same justification holds in the case that the value of $k$ is already
determined before running the algorithm, like in virtual database search;
this situation is indistinguishable from the random generation of $k$ at the
end of the algorithm, since state vector reduction can be backdated so that $%
k$ is already determined before running the algorithm. With $k$
predetermined, the preparation of register $K$ in an even superposition of
all the possible values of $k$ represents the initial ignorance of the value
of $k$ on the part of the second player. Since there is no more
determination of the column on the part of the second player, mutual
determination between the contents of registers $K$ and $X$ becomes
"precognition" of the column on the part of the second player.
"Precognition" corresponds to backdating, to before running the algorithm,
the state vector reduction induced by the measurement of (say) the second
bit of register $X$, which leaves (as said before) the initial preparation
of register $X$ unaltered and determines the second bit in the initial
preparation of register $K\ $(determines the column), reducing the initial
ignorance of the second player about the value of $k$. The related
information gain is

\begin{equation}
\Delta S=\frac{1}{2}\lg N,  \label{lg}
\end{equation}%
one bit with $N=4$. Besides database size, $N$ is the ratio between the size
of the superposition before measurement (8 terms with amplitudes even in
modulus -- see eq. \ref{gstate}) and the size of the subspace on which the
superposition is projected by quantum measurement (the 2 dimensions of the
Hilbert space of register $F$).

I should like to quote the question raised by Grover in his 2001 paper: "%
\textit{What is the reason that one would expect that a quantum mechanical
scheme could accomplish the search in }$O\left( \sqrt{N}\right) $\textit{\
steps? It would be insightful to have a simple two line argument for this
without having to describe the details of the search algorithm.}" The
"precognition" explanation might provide this argument. Casting it in two
lines: "the speed up is the reduction of the initial ignorance of the
solution due to backdating, to before running the algorithm, a
time-symmetric part of the state vector reduction on the solution".

A similar extension of Deutsch's algorithm yields the state before
measurement:

\begin{equation}
\frac{1}{2\sqrt{2}}\left[ \left( \left\vert 00\right\rangle _{K}+\left\vert
11\right\rangle _{K})\left\vert 0\right\rangle _{X}+(\left\vert
01\right\rangle _{K}+\left\vert 10\right\rangle _{K})\left\vert
1\right\rangle _{X}\right) \right] \left( \left\vert 0\right\rangle
_{F}-\left\vert 1\right\rangle _{F}\right)  \label{dstate}
\end{equation}%
where $k=00,~01,~10,~11\ $specifies the function randomly chosen by the
first player and $x$ is the answer provided by the second player (whether
the function is balanced or constant). State (\ref{dstate}) is reached by
invoking the computation of the function only once instead of the two times
required in the classical case. Although things are less symmetrical than in
database search, as the two registers have different length, there is still
a succession of entanglement and disentanglement, and mutual determination
between the contents of the two registers $K$ and $X$. The information gain $%
\Delta S=\frac{1}{2}\lg N$,$~$associated to backdating a time symmetric part
of the state vector reduction on the solution, is one bit -- the ratio of
Hilbert space sizes before and after measurement is still $N=4$. This is
consistent with the fact that the speed up of Deutsch's algorithm consists
in having to check the value of one bit rather than two in the classical
case.

Equation (\ref{lg}) can be rewritten by noting that $\lg N$, the logarithm
of the squeeze of Hilbert space size, is the von Neumann entropy of the
reduced density operator of register $K$ in state (\ref{gstate}):

\begin{equation}
\Delta S=\frac{1}{2}\Delta R.  \label{onehalf}
\end{equation}%
I call this entropy $\Delta R$ since it is also the decrease of entropy of
register $K$ during state vector reduction -- before reduction $K$ is
maximally entangled, after reduction it is in a sharp state and its entropy
is zero. $\Delta R$ can be used as an entropic measure of state vector
reduction. It is more general than the logarithm of the ratio of Hilbert
space sizes, with which it coincides in the case of even modulus amplitudes.

Similarly, the information gain $\Delta S$ associated with partial backdated
state vector reduction can be used as a measure of the speed up. This means
defining the speed up -- when applicable -- as the reduction of the
logarithmic size of the problem such that: \textit{the time taken by the
quantum algorithm to solve the problem is the same as the time taken by the
classical algorithm to solve the reduced problem.}

For example, in the case of Grover's algorithm, if database size is $N=4$,
the logarithmic size of the problem is $\lg 4=2$ (the number of bits of
register $K$), the logarithmic size of the reduced problem is $\lg 2=1$. The
time taken by the quantum algorithm to solve the problem of $2$ bits is the
same as the time taken by the classical algorithm to solve the problem of $1$
bit -- in both cases $\delta \left( k,x\right) $ is computed once.

Equation (\ref{onehalf}) states that this measure of the speed up is 50\% of
the entropic measure of state vector reduction in both Grover's \ and
Deutsch's algorithms. These algorithms concern unstructured problems. More
in general, the notion that the speed up is partial backdated state vector
reduction implies:

\begin{equation}
\Delta S\leq \Delta R,  \label{inequality}
\end{equation}%
where $\Delta R$, the entropic measure of state vector reduction, can be
defined in general as the entropy of the reduced density operator of the
observable being measured. We can do without the details of the quantum
algorithm by considering the state immediately before the measurement
projection, when the observable is maximally entangled with the pointer of
the measurement apparatus. In particular, inequality (\ref{inequality})
states that, when the problem-solution interdependence is physically
represented, there is no speed up without state vector reduction.

\section{Conclusions}

I would like to conclude by positioning simultaneous computation within the
development of quantum computation. It took a fundamental computation model
to describe computation in the quantum framework (Bennett, 1973, 1982,
Fredkin and Toffoli, 1982). It was naturally the model of deterministic
reversibility. The early works of Benioff (1982) and Feynman (1985) were
quantum mechanical models of deterministic reversible computation. I argue
that all quantum algorithms, starting with the first work of Deutsch (1985),
call for the extension of the early notion of deterministic reversible
computation to simultaneous computation. Summing up, this extension can be
segmented into the following steps: (i) replace the idealized
(deterministic, two body) bounching ball model of reversible computation by
the idealized (nondeterministic, many body) classical model of simultaneous
computation, (ii) notice that the simultaneous dependence between the
coordinates of the machine parts can be replaced, in the quantum framework,
by the simultaneous dependence between the populations of the reduced
density operators of the parts of the computer register, (iii) complete the
physical representation of problem-solving by comprising the
problem-solution interdependence, and (iv) see that the speed up is the gain
of information due to backdating, to before running the algorithm, a time
symmetric part of the state vector reduction on the solution.

\begin{acknowledgement}
Thanks are due to David Finkelstein and Artur Ekert for encouragement to
write down my way of thinking about quantum computation and, extended to
Shlomit Ritz Finkelstein, for stimulating discussions.
\end{acknowledgement}

\textbf{References}

Benioff, P. (1982). Quantum mechanical Hamiltonian models of Turing machines.%
\textit{\ J. Stat. Phys}., 29, 515.

Bennett, C.H. (1973). Logical reversibility of computation.\textit{\ IBM J.
Res. Dev.} 6, 525.

Bennett, C.H. (1982). The Thermodynamics of Computation -- a Review. \textit{%
Int. J. Theor. Phys.} 21, pp. 905.

Castagnoli, G. (1999). A quantum logic gate representation of quantum
measurement: reversing and unifying the two steps of von Neumann's model.
quant-ph/9912020.

Castagnoli, G., Monti, D., and Sergienko, A. (1999). Performing quantum
measurement in suitably entangled states originates the quantum computation
speed up. quant-ph/9908015.

Castagnoli, G., and Finkelstein, D. (2001). Theory of the quantum speed up. 
\textit{Proc. Roy. Soc. Lond}. A 457, 1799. quant-ph/0010081.

Cleve, R., Ekert, A., Macchiavello, C., and Mosca, M. (1997). Quantum
Algorithms Revisited. \textit{Phil. Trans. R. Soc. Lond.} A,
quant-ph/9708016.

Deutsch, D. (1985). Quantum theory, the Church-Turing principle and the
universal quantum computer. \textit{Proc. Roy. Soc}. (Lond.) A, 400, 97.

Feynman, R. P. (1985). Quantum mechanical Computers. Opt. News 11, 11;
reprinted in \textit{Found. Phys.} V 16, N. 6 (1986).

Finkelstein, D. R. (2007). Generic Relativity. Preprint.

Fredkin, E. and Toffoli, T. (1982). Conservative logic.\textit{\ Int. J.
Theor. Phys.} 21, 219.

Grover, L. K. (1996). A fast quantum mechanical algorithm for database
search. Proc. 28th Ann. ACM Symp. Theory of Computing.

Grover, L. K. (2001). From Schrodinger Equation to Quantum Search Algorithm.
Quant-ph/0109116.

Mulligan, K. and Smith, B. (1988). Mach and Ehrenfels: The Foundations of
Gestalt Theory. From Barry Smith (ed.), Foundations of Gestalt Theory,
Munich and Vienna: Philosophia, 124.
http://ontology.buffalo.edu/smith/articles/mach/mach.html - N\_1\_

Raussendorf, R. and Briegel, H. J. (2000). Quantum computing via
measurements only. arXiv:quant-ph/0010033 v1 7 Oct 2000.

Shor, P. (1994). Algorithms for quantum computation: Discrete logarithms and
factoring. \textit{Proc. 35th Ann. Symp. on Foundations of Comp. Sci.},
124--134.

Simon, D. (1994). On the Power of Quantum Computation. \textit{Proc. 35th
Ann. Symp. on Foundations of Comp. Sci.} 116--123.

\end{document}